\pdfoutput=1
\documentclass[%
reprint,
 amsmath,amssymb,
 aps,
floatfix,
]{revtex4-1}

\usepackage{graphicx}
\usepackage{dcolumn}
\usepackage{bm}
\usepackage{hyperref}
\usepackage{multirow}
\usepackage{amsmath,accents} 

\RequirePackage{xspace}

\RequirePackage{xspace}

\newcommand{\SFO}{SrFe$_{12}$O$_{19}$}
\newcommand{\SFAO}{SrFe$_{12-x}$Al$_{x}$O$_{19}$}
\newcommand{\exptval}[1]{\left<{#1}\right>}

\newcolumntype{.}{D{.}{.}{-1}}

\begin{document}

\title{Site occupancy and magnetic properties of Al-substituted
  M-type strontium hexaferrite}

\author{Vivek~Dixit}
\author{Chandani~N.~Nandadasa}
\affiliation{%
  Department of Physics and Astronomy, 
  Mississippi State University,
  Mississippi State, MS 39762, USA
}
\affiliation{%
  Center for Computational Sciences, 
  Mississippi State University, 
  Mississippi State, MS 39762, USA
}

\author{Sungho~Kim}
\affiliation{%
  Center for Computational Sciences, 
  Mississippi State University, 
  Mississippi State, MS 39762, USA
}

\author{Seong-Gon~Kim}
\email[Corresponding author: ]{kimsg@ccs.msstate.edu}
\affiliation{%
  Department of Physics and Astronomy, 
  Mississippi State University,
  Mississippi State, MS 39762, USA
}
\affiliation{%
  Center for Computational Sciences, 
  Mississippi State University, 
  Mississippi State, MS 39762, USA
}

\author{Jihoon~Park}
\author{Yang-Ki~Hong}
\affiliation{%
  Department of Electrical and Computer Engineering and MINT Center, 
  The University of Alabama,
  Tuscaloosa, AL 35487, USA
}

\author{Laalitha~S.~I.~Liyanage}
\affiliation{%
  Department of Physics, 
  University of North Texas,
  Denton, TX 76203, USA
}

\author{Amitava~Moitra}
\affiliation{%
  Thematic Unit of Excellence on Computational Materials Science,
  S.N. Bose National Centre for Basic Sciences, 
  Sector-III, Block-JD, Salt Lake, 
  Kolkata-700098, India
}

\date{April 6, 2015}

\begin{abstract}
  We use first-principles total-energy calculations based on density
  functional theory to study the site occupancy and magnetic
  properties of Al-substituted $M$-type strontium hexaferrite
  SrFe$_{12-x}$Al$_{x}$O$_{19}$ with $x=0.5$ and $x=1.0$.  We find
  that the non-magnetic Al$^{3+}$ ions preferentially replace
  Fe$^{3+}$ ions at two of the majority spin sites, $2a$ and $12k$,
  eliminating their positive contribution to the total magnetization
  causing the saturation magnetization $M_s$ to be reduced as Al
  concentration $x$ is increased.  Our formation probability analysis
  further provides the explanation for increased magnetic anisotropy
  field when the fraction of Al is increased.  Although Al$^{3+}$ ions
  preferentially occupy the $2a$ sites at a low temperature, the
  occupation probability of the $12k$ site increases with the rise of
  the temperature.  At a typical annealing temperature ($>
  700\,^{\circ}{\rm C}$) Al$^{3+}$ ions are much more likely to occupy
  the $12k$ site than the $2a$ site.  Although this causes the
  magnetocrystalline anisotropy $K_1$ to be reduced slightly, the
  reduction in $M_s$ is much more significant.  Their combined effect
  causes the anisotropy field $H_a$ to increase as the fraction of Al
  is increased, consistent with recent experimental measurements.
\end{abstract}


\maketitle

\section{Introduction}
\label{sec:intro} 

Strontium hexaferrite, \SFO\ (SFO) is one of the most commonly used
materials for permanent magnets, magnetic recording and data storage,
and components in electrical devices operating at microwave/GHz
frequencies, due to its high Curie temperature, large saturation
magnetization, excellent chemical stability and low manufacturing cost
\cite{Wang2012, Pullar2012, Pang2010, Davoodi2011, Ashiq2012}.
However, in comparison with Nd-Fe-B and magnet, the coercivity of the
SFO is low and presents a significant limitation in its application.
Therefore, enhancement of the coercivity is an important research
topic for the strontium hexaferrite.

In order to tailor the magnetic properties such as magnetization and
coercivity, various cation substitutions in the M-type hexaferrites
have been investigated.  For example, the substitution of La
\cite{Seifert2009, Wang2004}, Sm \cite{Wang2001}, Pr \cite{Wang2005}
and Nd \cite{Wang2002} in the SFO increased coercivity moderately
while the substitution of Zn-Nb \cite{Fang2004}, Zn-Sn
\cite{Ghasemi2010, Ghasemi2011, Liyanage2013} and Sn-Mg
\cite{Davoodi2011} decreased coercivity.  However, the coercivity of
the M-type hexaferrites is not increased significantly by these cation
substitutions, and is still much smaller than that of Nd-Fe-B magnet
\cite{Liu2010}.

Al substitution in the M-type hexaferrite has been more effective in
enhancing coercivity \cite{Bertaut1959, Wang2000, Liu2011, Luo2012,
  Harward2013}.  Particularly, Wang et al synthesized Al-doped SFO
\SFAO\ (SFAO) with Al content of $x = 0 - 4$ using glycin–nitrate
method and subsequent annealing in a temperature over
$700\,^{\circ}{\rm C}$ obtaining the largest coercivity of 17.570~kOe,
which is much larger than that of SFO (5.356~kOe) and exceeds even the
coercivity of the Nd$_2$Fe$_{17}$B (15.072~kOe) \cite{Wang2012}.  Wang
and co-workers also observed that the coercivity of the SFAO increases
with increasing Al concentration at a fixed annealing temperature.
These results call for a systematic understanding, from first
principles, of why certain combinations of dopants lead to particular
results. This theoretical understanding will be essential in
systematically tailoring the properties of SFO.

There have been several previous first-principles investigations of
SFO.  Fang et al investigated the electronic structure of SFO using
density-functional theory (DFT) \cite{Fang2003}.  Park et al have
calculated the exchange interaction of SFO from the differences of the
total energy of different collinear spin configurations
\cite{Park2014}.  In spite of the importance of substituted SFO, only
a few theoretical investigations have been done.  Magnetism in La
substituted SFO has been studied using DFT \cite{Kupferling2005,
  Novak2005a}.  The site occupancy and magnetic properties of Zn-Sn
substituted SFO has been investigated \cite{Liyanage2013}.

In this work we use first-principles total-energy calculations to
study the site occupation and magnetic properties of Al substituted
$M$-type strontium hexaferrite \SFAO\ with $x=0.5$ and $x=1.0$.  Based
on DFT calculations, we determine the the structure of various
configurations of SFAO with different Al concentrations and compute
the occupation probabilities for different substitution sites at
elevated temperatures.  We show that our model predicts an decrease of
saturation magnetization as well as a decrease in magnetocrystalline
anisotropy $K_1$, and the increase of the anisotropy field $H_a$ as
the fraction of Al is increased, consistent with recent experimental
measurements.

\section{Methods}
\label{sec:method}

\begin{figure}[tpb]
  \centering
  \includegraphics[scale=0.5,keepaspectratio=true]{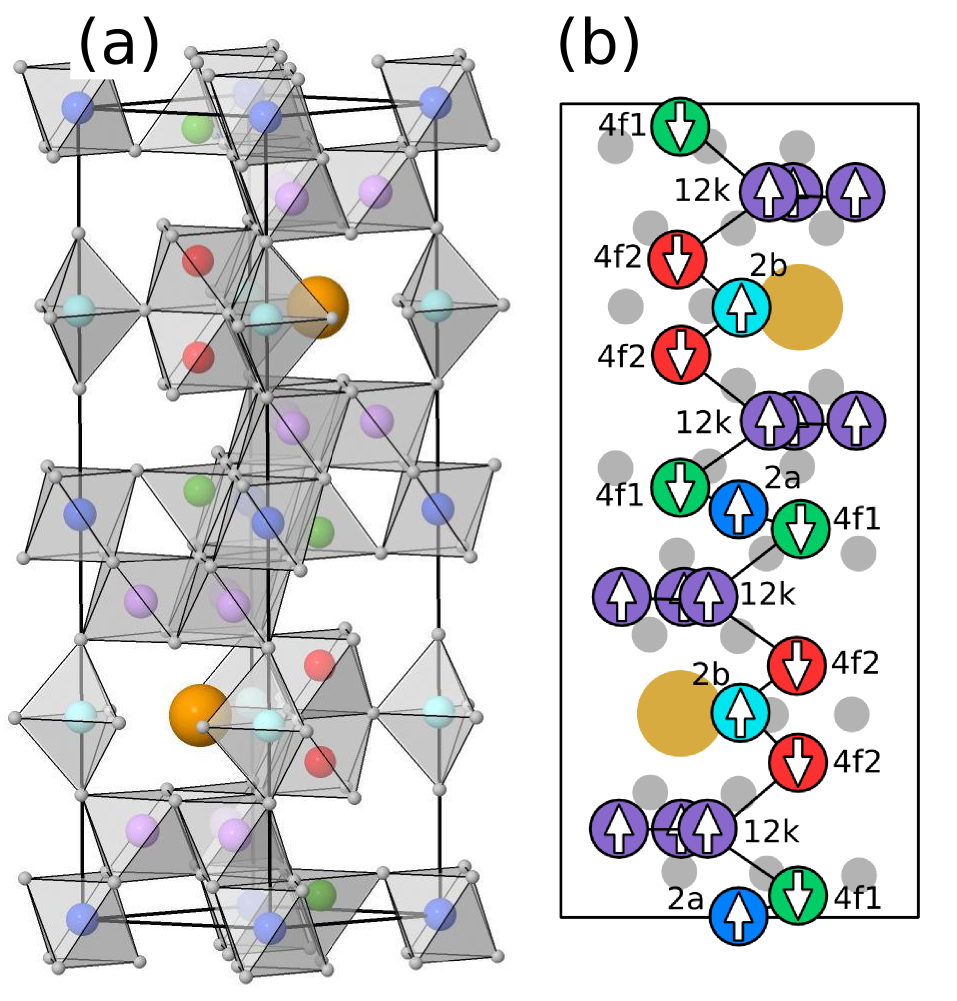}
  \caption{\label{fig:SFO-Fe-sites} (color online) (a) A unit cell of
    SFO containing two formula units. Two large gold spheres are Sr
    atoms and small gray spheres are O atoms.  Colored spheres
    enclosed by polyhedra formed by O atoms represent Fe$^{3+}$ ions
    in different inequivalent sites: $2a$ (blue), $2b$ (cyan), $12k$
    (purple), $4f_1$ (green), and $4f_2$ (red). (b) A schematic
    diagram of the lowest-energy spin configuration of Fe$^{3+}$ ions
    of SFO.  The arrows represent the local magnetic moment at each
    atomic site. (For interpretation of the references to color in
    this figure caption, the reader is referred to the online version
    of this paper.)}
\end{figure} 

SFO has a hexagonal magnetoplumbite crystal structure that belongs to
$P63/mmc$ space group. Fig.~\ref{fig:SFO-Fe-sites} shows a unit cell
of SFO used in the present work that contains 64 atoms of two formula
units.  Magnetism in SFO arises from Fe$^{3+}$ ions occupying five
crystallographically inequivalent sites in the unit cell, three
octahedral sites ($2a$, $12k$, and $4f_{2}$), one tetrahedral site
($4f_{1}$), and one trigonal bipyramidal site ($2b$) as represented by
the polyhedra in Fig.~\ref{fig:SFO-Fe-sites}(a). SFO is also a
ferrimagnetic material that has 16 Fe$^{3+}$ ions with spins in the
majority direction ($2a$, $2b$, and $12k$ sites) and 8 Fe$^{3+}$ ions
with spins in the minority direction ($4f_1$ and $4f_2$ sites) as
indicated by the arrows in Fig.~\ref{fig:SFO-Fe-sites}(b).

Total energies and forces were calculated using DFT with projector
augmented wave (PAW) potentials as implemented in VASP
\cite{Kresse1996, Kresse1999}.  All calculations were spin polarized
according to the ferrimagnetic ordering of Fe spins as first proposed
by Gorter \cite{Fang2003, Gorter1957}.  A plane-wave energy cutoff of
520~eV was used both for pure SFO and Al-substituted SFO.  Reciprocal
space was sampled with a $7\times 7\times 1$ Monkhorst-Pack mesh
\cite{Monkhorst1976} with a Fermi-level smearing of 0.2~eV applied
through the Methfessel-Paxton method \cite{Methfessel1989}.  We
performed relaxation of the electronic degrees of freedom until the
change in free energy and the band structure energy was less than
$10^{-7}$~eV.  We performed geometric optimizations to relax the
positions of ions and cell shape until the change in total energy
between two ionic step was less than $10^{-4}$~eV.  Electron exchange
and correlation was treated with the generalized gradient
approximation (GGA) as parameterized by the Perdew-Burke-Ernzerhof
(PBE) scheme \cite{Perdew1996}.  To improve the description of
localized Fe $3d$ electrons, we employed the GGA+U method in the
simplified rotationally invariant approach described by Dudarev et al
\cite{Dudarev1998}.  This method requires an effective $U$ value
($U_{\mathrm{eff}}$) equal to the difference between the Hubbard
parameter $U$ and the exchange parameter $J$.  We chose
$U_{\mathrm{eff}}$ equal to 3.7~eV for Fe based on the previous result
\cite{Liyanage2013}.

\section{Results and Discussion}
\label{sec:results}

\begin{figure}[tbp]
  \centering
  \includegraphics[scale=0.35,keepaspectratio=true]{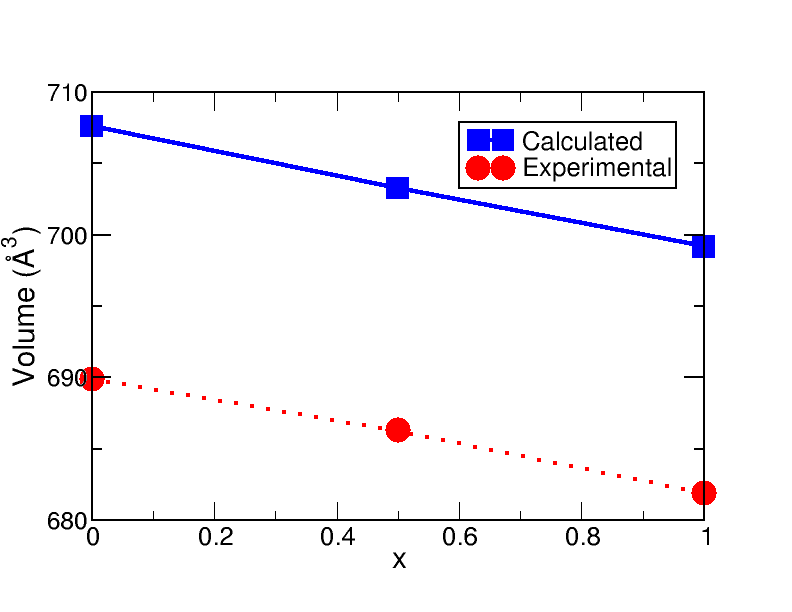}
  \caption{\label{fig:volume} (color online) Comparison of calculated
    and experimental (Ref.~[\onlinecite{Luo2012}]) volume of the unit
    cell of \SFAO\ as a function of the fraction of Al $x$.}
\end{figure} 
The substitution of Fe$^{3+}$ ions by Al$^{3+}$ ions considerably
affects the unit cell parameters.  We have calculated the lattice
parameters of pure and Al-substituted SFO by relaxing ionic positions
as well as the volume and shape of the unit cell.  In all cases the
final unit cell was found to remain hexagonal.  In the case of pure
SFO, the lattice parameters $a$ and $c$ were found to be 5.93~\AA\ and
23.21~\AA\ in good agreement with the experimental values of
$a=5.88$~\AA\ and $c=23.04$~\AA, respectively \cite{Luo2012,
  Kimura1990}; the deviation between the experimental and the
theoretical values is less than 1\%. In the case of $x=0.5$ in \SFAO\,
the lattice parameters $a$ and $c$ were calculated to be $5.92$~\AA\
and $23.16$~\AA\, respectively, while the volume of the unit cell was
reduced by 0.61\%. For $x=1.0$, $a=5.91$~\AA\ and $c=23.04$~\AA\ were
found, and reduction in the unit cell volume was 2.51\%.
Fig.~\ref{fig:volume} shows that the reduction of unit cell volume
predicted by our DFT calculation is consistent with the experimental
results \cite{Wang2012, Luo2012}.

\begin{figure}[tbp]
  \centering
  \includegraphics[scale=0.57,keepaspectratio=true]{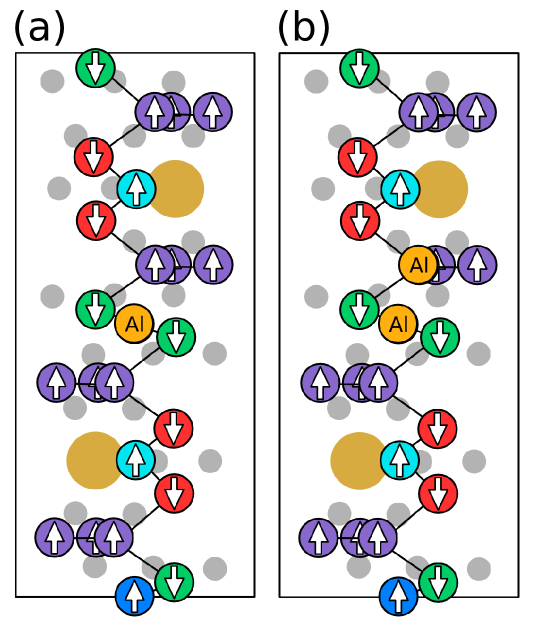}
  \caption{\label{fig:SFAO-configs} (color online) The structures of
    \SFAO\ with spins oriented in the easy axis (001): (a)
    configuration $[2a]$ for $x=0.5$ and (b) configuration
    $[2a,12k].1$ for $x=1.0$.  Al atoms are labeled and other atoms
    are colored as in Fig.~\ref{fig:SFO-Fe-sites}.  }
\end{figure} 
We investigated the site preference of Al substituting Fe in \SFAO\
for (i) $x=0.5$ and, (ii) $x=1.0$.  The $x=0.5$ case corresponds to
the condition where one Al atom is substituted in the unit cell, while
two Al atoms were substituted in the case of $x=1.0$ as shown
Fig.~\ref{fig:SFAO-configs}.  To determine the site preference of the
substituted Al atoms, the substitution energy of configuration $i$ was
calculated using the following expression:
\begin{equation} \label{eq:Esub}
  E_{\text{sub}}(i) = E(\text{SFAO}(i)) -E(\text{SFO}) 
  -\sum_{\alpha}n_{\alpha}\epsilon(\alpha) 
\end{equation}
where $E(\text{SFAO}(i))$ is the total energy per unit cell at 0~K for
SFAO in configuration $i$ while $E(\text{SFO})$ is the total energy
per unit cell at 0~K for SFO.  $\epsilon(\alpha)$ is the total energy
per atom for element $\alpha$ ($\alpha$ = Al, Fe) at 0~K in its most
stable crystal structure. $n_\alpha$ is the number of atoms of type
$\alpha$ added: if two atoms are added then $n_{\alpha} = +2$ while
$n_\alpha = -1$ when one atom is removed. The configuration with the
lowest $E_{\text{sub}}$ is concluded to be the ground state
configuration, and the corresponding substitution site is the preferred
site of Al atoms at 0~K.

To understand the site preference of the substituted Al$^{3+}$ ions at
higher temperatures, we compute the formation probability of
configuration $i$ using the Maxwell-Boltzmann statistical distribution
\cite{Liu:2014}:
\begin{equation} \label{eq:P(i)} 
  P_{i} = \frac {g_{i}\exp(-\Delta G_{i}/k_{B}T)}
  {\sum_{j} g_{j}\exp(-\Delta G_{j}/k_{B}T)}
\end{equation} 
where $g_{i}$ is the multiplicity of configuration $i$ (number of
equivalent configurations) and
\begin{equation} \label{eq:G(i)}
  \Delta G_{i} = \Delta E_{\text{sub}}(i) + P\Delta V_{i} - T\Delta S_{i}   
\end{equation}
is the change of the free energy of configuration $i$ relative to that
of the ground state configuration; $\Delta E_{\text{sub}}(i)$, $\Delta
V_{i}$, and $\Delta S_{i}$ are the substitution energy change, volume
change, entropy change for configuration $i$; $P$ and $k_{\text{B}}$
are the pressure and Boltzmann constant.

For the $x=0.5$ concentration, one Al atom is substituted at one of
the 24 Fe sites of the unit cell as shown in
Fig.~\ref{fig:SFAO-configs}(a).  The application of crystallographic
symmetry operations shows that many of these Fe sites are equivalent
and leaves only five inequivalent structures.  We label these
inequivalent configurations using the crystallographic name of the
Fe site: $[2a]$, $[2b]$, $[4f_{1}]$, $[4f_{2}]$, and $[12k]$.  These
structures were created by substituting one Al atom to the respective
Fe site of a SFO unit cell and performing full optimization of the
unit cell shape and volume, and ionic positions.

\begin{table}[tbp]
  \caption{\label{tab:SFAOx0.5} Five inequivalent configurations of \SFAO\ 
    with $x=0.5$.  $g$ is the multiplicity of the configuration.  
    $E_{\text{sub}}$ is the substitution energy of the SFAO. 
    The total magnetic moment ($m_{\text{tot}}$) and its change with respect 
    to SFO ($\Delta m_{\text{tot}}$) are also given. 
    All values are for a double formula unit cell containing 64 atoms. 
    Energies are in eV while magnetic moments are in $\mu_{\text{B}}$.
  }
  \begin{ruledtabular}
    \begin{tabular}{ccccc}
      Config & $g$ &  $E_{\text{sub}}$ & $m_{\text{tot}}$ & $\Delta m_{\text{tot}}$ \\
      \hline
      $[2a]$    & 2   &-6.04 & 35 & -5 \\
      $[12k]$   & 12  & -6.00 & 35 & -5 \\
      $[4f_{2}]$ & 4  & -5.63 & 45 & +5 \\
      $[2b]$    & 2   & -5.60 & 35 & -5 \\
      $[4f_{1}]$ & 4  & -5.57 & 45 & +5 \\
    \end{tabular}
  \end{ruledtabular}
\end{table}

Table~\ref{tab:SFAOx0.5} lists the results of our calculation for all
five inequivalent configurations in the order of increasing
substitution energy.  The lowest $E_{\text{sub}}$ is found for
configuration $[2a]$ shown in Fig.~\ref{fig:SFO-Fe-sites}(a).  We can
conclude that at 0~K the most preferred site for the substituted Al
atom is the $2a$ site.  We used Eq.~(\ref{eq:P(i)}) to compute the
probability to form each configuration as a function of temperature.
Since the volume change among different configurations is very small
(less than 0.1~$\AA^3$), we can safely regard $P\Delta V$ term to be
negligible (in the order of $10^{-7}$~eV at the standard pressure of
1~atm) compared to the $\Delta E_{\text{sub}}(i)$ term in
Eq.~(\ref{eq:G(i)}).  The entropy change $\Delta S$ has a
configurational part, $\Delta S_{c}$, and a vibrational part, $\Delta
S_{\text{vib}}$ \cite{vandeWalle2002}.  For binary substitutional
alloys such as the present system, $\Delta S_{\text{vib}}$ is around
0.1-0.2~$k_{\text{B}}$/atom, and $\Delta S_{c}$ is
0.1732~$k_{\text{B}}$/atom \cite{Liu:2014}. Therefore, we set $\Delta
S = 0.3732$~$k_{\text{B}}$/atom.

\begin{figure}[tbp]
  \centering
  \includegraphics[scale=0.33,keepaspectratio=true]{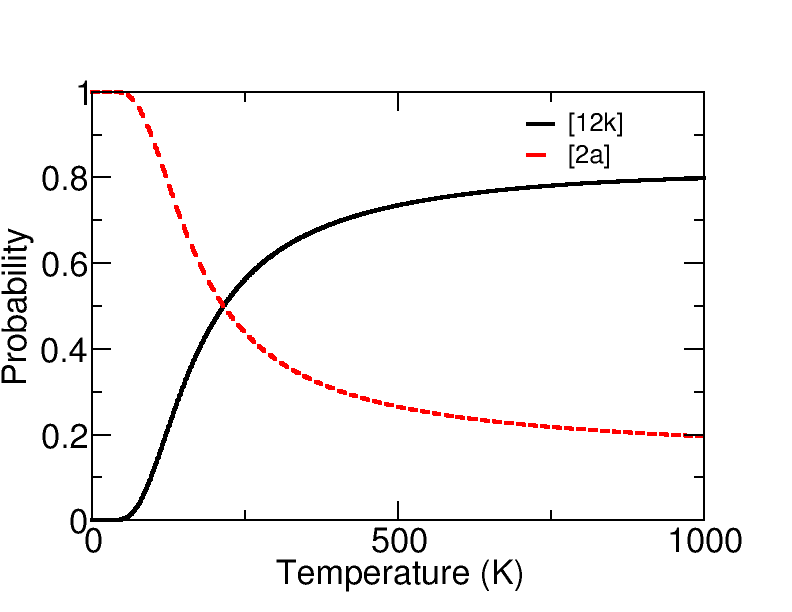}
  \caption{\label{fig:Prob:Al1} (color online) Temperature dependence
    of the formation probability of different configurations of \SFAO\
    with $x=0.5$.  The configurations with negligible probability are
    not shown.}
\end{figure} 

Fig.~\ref{fig:Prob:Al1} displays the temperature dependence of the
formation probability of different configurations of \SFAO\ with
$x=0.5$.  The doped Al$^{3+}$ ions mainly replace Fe$^{3+}$ ions from
the $2a$ and the $12k$ sites. The formation probabilities of $[2b]$,
$[4f_{1}]$ and $[4f_{2}]$ are negligible and not shown in
Fig.~\ref{fig:Prob:Al1}.  The probability that the doped Al$^{3+}$ ion
replaces Fe$^{3+}$ ion from the $2a$ site is maximum at 0~K and it
falls as temperature increases, while the occupancy of Al$^{3+}$ at
the $12k$ site rises with temperature.  The two curves cross at $T
\sim 220$~K.  At a typical annealing temperature of 1000~K for SFAO
\cite{Wang2012} the site occupation probability of the site $2a$ and
$12k$ is 0.196 and 0.798, respectively.  Thus, during the annealing
process of the synthesis of the SFAO the doped Al$^{3+}$ ions are more
likely to replace Fe$^{3+}$ ions from the $12k$ site than the $2a$
site despite of higher substitution energy.

For the $x=1.0$ concentration, two Al atoms are substituted at two of
the 24 Fe sites of the unit cell as shown in
Fig.~\ref{fig:SFAO-configs}(b).  These Fe sites have more than one
equivalent site.  Substitution of Al atoms breaks the symmetry of the
equivalent sites of pure SFO.  Out of all $C(24,2)=276$ possible
structures, many of the structures are crystallographically
equivalent.  On applying crystallographic symmetry operations, the
number of inequivalent structures reduces to 40.  We label these
inequivalent configurations using the convention of [(site for the
first Al),(site for the 2nd Al)].(unique index).  For example, when
two Al atoms are substituted at the $2a$ and $12k$ sites, there are 2
inequivalent configurations, which are labeled as $[2a,12k].1$ and
$[2a,12k].2$.  These structures are fully optimized and their
substitution energies are calculated using Eq.~(\ref{eq:Esub}).
When there are more than one inequivalent configuration, we assign the
unique index in the order of increasing $E_{\text{sub}}$.

\begin{table}[tbp]
  \caption{\label{tab:SFAOx1.0} Ten lowest energy inequivalent 
    configurations of \SFAO\ with $x=1.0$.  
    $g$ is the multiplicity of the configuration.
    $E_{\text{sub}}$ is the substitution energy per Al atom.  
    The total magnetic moment ($m_{\mathrm{tot}}$) and its change with respect 
    to SFO ($\Delta m_{\mathrm{tot}}$) are also given. 
    All values are for a double formula unit cell containing 64 atoms. 
    Energies are in eV while moments are in $\mu_{\text{B}}$.}
  \begin{ruledtabular}
    \begin{tabular}{ccccc}
      Config & $g$ &  $E_{\text{sub}}$ & $m_{\text{tot}}$ & $\Delta m_{\text{tot}}$ \\
      \hline
      $[2a,2a]$      & 1   & -6.056 & 30 & -10 \\
      $[2a,12k].1$   & 12  & -6.054 & 30 & -10 \\
      $[2a,12k].2$   & 12  & -6.041 & 30 & -10 \\
      $[12k,12k].1$  & 6   & -6.025 & 30 & -10 \\
      $[12k,12k].2$  & 12  & -6.025 & 30 & -10 \\
      $[12k,12k].3$  & 12  & -6.027 & 30 & -10 \\
      $[12k,12k].4$  & 12  & -6.025 & 30 & -10 \\
      $[12k,12k].5$  & 6   & -6.023 & 30 & -10 \\
      $[12k,12k].6$  & 6   & -6.017 & 30 & -10 \\
      $[12k,12k].7$  & 12  & -6.014 & 30 & -10 \\
    \end{tabular}
  \end{ruledtabular}
\end{table}
\begin{figure}[tbp]
  \centering
  \includegraphics[scale=0.33,keepaspectratio=true]{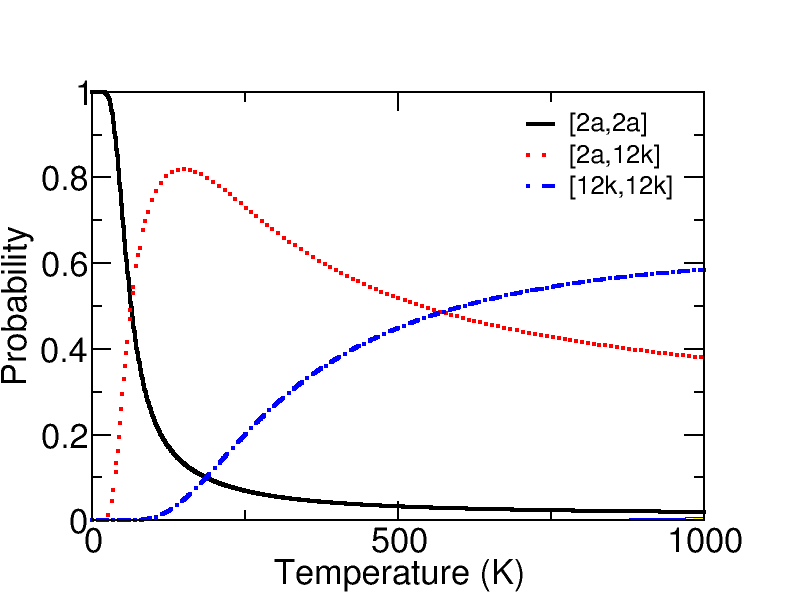}
  \caption{\label{fig:Prob:Al2} (color online) Temperature dependence
    of the formation probability of different configurations of \SFAO\
    with $x=1.0$. For clarity only the configurations with significant
    formation probability are labeled.}
\end{figure} 

Table~\ref{tab:SFAOx1.0} lists the ten lowest energy configurations of
\SFAO\ with $x=1.0$. The configuration $[2a,2a]$ where two Al$^{3+}$
ions replace Fe$^{3+}$ ions from two $2a$ sites has the lowest
$E_{\text{sub}}$, and it is the most energetically favorable
configuration at 0~K. To investigate the site occupation at nonzero
temperatures we compute the formation probability of each
configuration using Eq.~(\ref{eq:P(i)}).  Similar to the previous case
the volume change among different configurations is very small (less
than 0.7~$\AA^3$) and we can safely ignore the $P\Delta V$ term.  The
entropy term is calculated in the same way as the $x=0.5$ case.
Fig.~\ref{fig:Prob:Al2} shows the variation of the formation
probability of different configurations with temperature.  We note
that due to low multiplicity of the configuration $[2a,2a]$, its
formation probability falls rapidly as temperature increases.  On the
other hand, the formation probability of the configuration $[2a,12k]$
(sum of the formation probabilities for all $[2a,12k].n$
configurations) increases steeply and reaches a maximum value at 50~K
and then falls with temperature.  Fig.~\ref{fig:Prob:Al2} shows that
the formation probability of the $[2a,12k]$ configuration becomes
larger that that of $[2a,2a]$ beyond $T \sim 10$~K, which is a much
lower transition temperature than in the $x=0.5$ case.

We can calculate the occupation probability of Al at nonzero
temperatures for a given site by adding all formation probabilities of
the configurations where at least one Al$^{3+}$ ion is substituted in
that site.  At the annealing temperature of 1000~K, the occupation
probability of Al for $12k$ site is 79.8\% for $x=0.5$ as given in
Table~\ref{tab:H_a}.  The same probability is increased to 97.7\% for
$x=1.0$ as calculated by adding the $P_{1000}$'s for all
configurations that contain the $12k$ site.  This means that the
fraction of Al$^{3+}$ ions occupying the $12k$ site increases when the
fraction of Al is increased from $x=0.5$ to $x=1.0$.  This conclusion
is in agreement with the previously reported measurements
\cite{Bertaut1959, Albanese1995421, Wang2012}.

\begin{table*}[tbp]
  \centering
  \caption{\label{tab:magmom-contrib} Contribution of atoms in each 
    sublattice to the total magnetic moment of Al-substituted SFO 
    structures $[12k]$, $[2a]$, and $[2a,12k].1$ compared with pure SFO. 
    All magnetic moments are in $\mu_{\text{B}}$. $\Delta m$ is 
    measured relative to the values for the pure SFO. 
    Note that the total magnetic moment of the unit cell ($m_{\mathrm{tot}}$) 
    is slightly different than the sum of local magnetic moments due to the 
    contribution from the interstitial region.} 
  \begin{ruledtabular}
    \begin{tabular}{c|rr|rrr|rrr|rrr}
      \multicolumn{1}{c|}{\multirow{2}{*}{site}} & \multicolumn{2}{c|}{SFO} 
      & \multicolumn{3}{c|}{$[12k]$} 
      & \multicolumn{3}{c|}{$[2a]$}
      & \multicolumn{3}{c}{$[2a,12k].1$} \\  
      \cline{2-12}
      & atoms & $m$ & atoms & $m$ & $\Delta m$ & atoms & $m$ & $\Delta m$ 
      & atoms & $m$ & $\Delta m$ \\
      \hline  
      $2d$ & 2 Sr & -0.006 & 2 Sr & -0.006 & 0.000 & 2 Sr & -0.006 & 0.000 
      & 2 Sr & -0.006 & 0.000 \\
      \hline
      \multirow{2}{*}{$2a$}      
      & 1 Fe &  4.156 & 1 Fe & 4.155 & -0.001 & 1 \textbf{Al} & -0.010 
      & -4.166 & 1 \textbf{Al} & -0.010 & -4.166 \\
      & 1 Fe &  4.156 & 1 Fe & 4.156 & 0.000 & 1 Fe & 4.156 & 0.000 
      & 1 Fe & 4.156 & 0.000 \\
      \hline
      $2b$ & 2 Fe & 8.098 & 2 Fe & 8.086 & -0.012 & 2 Fe & 8.100 & 0.001 
      & 2 Fe & 8.090 & -0.008 \\
      $4f_1$ & 4 Fe & -16.152 & 4 Fe & -16.189 & -0.037 & 4 Fe & -16.268 
      & -0.116 & 4 Fe & -16.304 & -0.152 \\
      $4f_2$ & 4 Fe & -16.384 & 4 Fe & -16.420 & -0.036 & 4 Fe & -16.382 
      & 0.002 & 4 Fe & -16.418 & -0.034 \\
      \hline
      \multirow{2}{*}{$12k$}
      & 1 Fe & 4.172 & 1 \textbf{Al} & 0.000 & -4.172 & 1 Fe & 4.170 
      & -0.002 & 1 \textbf{Al} & -0.001 & -4.173 \\
      & 11 Fe & 45.884 & 11 Fe & 45.861 & -0.023 & 11 Fe & 45.872 & -0.012 
      & 11 Fe & 45.846 & -0.038 \\
      \hline
      $4e$  & 4 O & 1.416 & 4 O & 1.304 & -0.112 & 4 O & 1.424 & 0.008 
      & 4 O & 1.316 & -0.100 \\
      $4f$ & 4 O & 0.360 & 4 O & 0.281 & -0.079 & 4 O & 0.310 & -0.050 
      & 4 O & 0.230 & -0.129 \\
      $6h$ & 6 O & 0.124 & 6 O & 0.115 & 0.009 & 6 O & 0.134 & 0.010 
      & 6 O & 0.117 & -0.007 \\
      $12k$ & 12 O & 1.016 & 12 O & 0.877 & -0.129 & 12 O & 0.548 & -0.468 
      & 12 O & 0.404 & -0.612 \\
      $12k$ & 12 O & 2.140 & 12 O & 1.895 & -0.245 & 12 O & 2.088 & -0.052 
      & 12 O & 1.839 & -0.301\\
      \hline
      $\sum m$ & & 38.980 & & 34.114 & -4.837 & & 34.136 & -4.845 & 
      & 29.259 & -9.720 \\
      $m_{\text{tot}}$ & & 40 & & 35 & -5 & & 35 & -5 & & 30 & -10 \\
    \end{tabular}
  \end{ruledtabular}
\end{table*}

In Table~\ref{tab:magmom-contrib} we compare the contribution of
different sublattices to the total magnetic moment in Al-substituted
SFO.  To see the effect of Al$^{3+}$ ions in different substitution
sites, we split the entries of sublattices containing these ions
($2a$ and $12k$).  As expected, Al$^{3+}$ ions carry negligible
magnetic moment regardless of their substitution sites.  Consequently,
when they replace Fe$^{3+}$ ions in the minority spin sites ($4f_1$
and $4f_2$), they eliminate a negative contribution and hence increase
the total magnetic moment.  On the other hand, when they replace
Fe$^{3+}$ ions in the majority spin sites ($12k$, $2a$, and $2b$),
they eliminate a positive contribution and hence reduce the total
magnetic moment.  For the $x=0.5$ case, the most probable sites are
$12k$ and $2a$ (majority sites) and the net magnetic moment of the
unit cell is reduced by 5~$\mu_{\text{B}}$. For the configuration
$[2a,12k].1$ of the $x=1.0$ case, two Al atoms are substituted in the
$2a$ and $12k$ sites, there is a reduction of 10~$\mu_{\text{B}}$ in
the total magnetic moment per unit cell.

\begin{table*}[tbp]
  \caption{\label{tab:H_a} The saturation magnetization ($M_s$), 
    magnetocrystalline anisotropy energy (MAE),  
    magnetocrystalline anisotropy constant ($K_1$) and anisotropy field 
    ($H_a$) for SFO and SFAO. 
    $x$ is the Al fraction in \SFAO\ and
    $V$ is the volume of the unit cell in \AA$^3$.
    $P_{1000}$ is the formation probability at 1000~K. 
    The averaged quantities are weighted by $P_{1000}$.  
    $M_s$ is in emu/g, MAE in meV, $H_a$ in kOe, and $K_1$ in kJ$\cdot$m$^{-3}$.
  }
  \begin{ruledtabular}
    \begin{tabular}{cc.ccc..ccc}
      $x$ & Config & \multicolumn{1}{c}{$M_s$} 
      & MAE & $V$ & $K_1$ & \multicolumn{1}{c}{$H_{a}$} 
      & \multicolumn{1}{c}{$P_{1000}$} 
      & $\exptval{M_s}$ 
      & $\exptval{K_1}$ & $\exptval{H_a}$ \\
      \hline
      0.0 & SFO & 110.19 & 0.85 & 707.29 & 193 & 7.35 & 1.000 
      & 110.19 & 193 & 7.35 \\ 
      \hline
      \multirow{5}{*}{  0.5 } & $[2a]$ & 96.41 & 0.95 & 703.29 & 216 & 9.38 
      & 0.196 & \multirow {5}{*}{96.49} 
      & \multirow {5}{*}{189} & \multirow {5}{*}{8.18} \\ 
      & $[12k]$ & 96.41 & 0.80 & 703.19 & 182 & 7.90 & 0.798 & \\
      & $[2b]$ & 96.41 & 0.67 & 702.82 & 152 & 6.62 & 0.003 & \\
      & $[4f_{1}]$ & 123.96 & 0.86 & 704.22 & 196 & 6.61 & 0.001 & \\
      & $[4f_{2}]$ & 123.96 & 0.83 & 702.58 & 189 & 6.38 & 0.001 & \\
      \hline
      \multirow{15}{*}{  1.0 }  & $[2a,2a]$ & 82.64 & 0.99 & 698.94 & 227 
      & 11.41 & 0.019 & \multirow {15}{*}{83.03} 
      & \multirow {15}{*}{184}& \multirow {15}{*}{9.23} \\
      & $[2a,12k]$ & 82.64 & 0.88 & 699.08 & 202 & 10.13 & 0.379 \\
      & $[12k,12k]$ & 82.64 & 0.75 & 698.66 & 172 & 8.64 & 0.585 \\
      & $[12k,4f_{2}]$ & 110.19 & 0.78 & 690.64 & 181 & 6.74 & 0.007 \\
      & $[12k,4f_{1}]$ & 110.19 & 0.80 & 700.29 & 183 & 6.92 & 0.004 \\
      & $[12k,2b]$ & 82.64 & 0.62 & 698.98 & 142 & 7.14 & 0.002 \\
      & $[4f_{2},4f_{2}]$ & 137.74 & 0.80 & 697.96 & 184 & 5.53 & 0.000 \\
      & $[4f_{2},4f_{1}]$ & 137.74 & 0.83 & 699.62 & 191 & 5.74 & 0.000 \\
      & $[4f_{2},2b]$ & 110.19 & 0.60 & 698.82 & 138 & 5.19 & 0.000 \\
      & $[4f_{2},2a]$ & 110.19 & 0.90 & 698.53 & 206 & 7.78 & 0.002 \\
      & $[4f_{1},4f_{1}]$ & 137.74 & 0.86 & 701.38 & 196 & 5.95 & 0.000 \\
      & $[4f_{1},2b]$ & 110.19 & 0.65 & 699.95 & 149 & 5.62 & 0.000 \\
      & $[4f_{1},2a]$ & 110.19 & 0.91 & 700.11 & 208 & 7.87 & 0.001 \\
      & $[2b,2b]$ & 82.64 & 0.45 & 698.86 & 103 & 5.19 & 0.000 \\
      & $[2b,2a]$ & 82.64 & 0.74 & 698.82 & 170 & 8.53 & 0.001 \\
    \end{tabular}
  \end{ruledtabular}
\end{table*}

Magnetic Anisotropy determines the capacity of a magnet to withstand
external magnetic and electric fields. This property is of
considerable practical interest, because anisotropy is exploited in
the design of the most magnetic materials of commercial
importance. The magnetocrystalline anisotropy energy (MAE) is one of
the main factors that determine the total magnetic anisotropy of the
material.  To investigate the effect of Al substitution on the
magnetic anisotropy of SFO, we computed the MAE and the magnetic
anisotropy constant of \SFAO\ for $x = 0, 0.5$ and $1$.  The MAE,
in the present case, is defined as the difference between the two
total energies where electron spins are aligned along two different
directions \cite{Ravindran:1999}:
\begin{equation}
  E_{\mathrm{MAE}} = E_{(100)} - E_{(001)}
\end{equation}
where $E_{(100)}$ is the total energy with spin quantization axis in
the magnetically hard plane and $E_{(001)}$ is the total energy with
spin quantization axis in the magnetically easy axis.  Using the MAE,
the uniaxial magnetic anisotropy constant $K_1$ can be computed
\cite{Munoz:2013, Smit:1959}
\begin{equation}
  K_1 = \frac{E_{\mathrm{MAE}}}{V\sin^2\theta}
\end{equation} 
where $V$ is the equilibrium volume of the unit cell and $\theta$ is
the angle between the two spin quantization axis orientations
(90$^\circ$ in the present case).  The anisotropy field $H_a$ can be
expressed as \cite{Kittel:1949}
\begin{equation}
  \label{eq:H_a}
  H_a = \frac{2K_1}{M_s}
\end{equation} 
where $K_1$ is a magnetocrystalline anisotropy constant and $M_s$ is
saturation magnetization.

The results for the MAE, the magnetocrystalline anisotropy constant
$K_1$, and anisotropy field $H_a$ for SFAO with different Al
concentration are presented in Table~\ref{tab:H_a}.  To compare with
experimental results, we also compute the weighted average of $K_1$
and $H_a$ using the formation probability $P_{1000}$ at a typical
annealing temperature of 1000~K \cite{Wang2012}.  We note that SFAO
considered in the present work loses most of its magnetic properties
at typical annealing temperatures (1000~K or higher) that are near or
above its Curie temperature.  The magnetic properties listed in
Table~\ref{tab:H_a} refer to their ground state properties at the
temperature $T=0$.  We use the formation probability at 1000~K to
compute the weighted averages as the crystalline configurations of
SFAO will be distributed according to this value during the annealing
process.

Table~\ref{tab:H_a} shows that $M_s$ decreases as the concentration of
Al $x$ is increased from 0 to 0.5 to 1.0, consistent with the previous
experimental results \cite{Albanese1979, Harward2013, Ustinov:2009,
  Wang2012}.  Our calculation also shows that $K_1$ decreases as the
concentration of Al $x$ is increased from 0 to 0.5 to 1.0.  At a low
temperature Al atoms prefer to occupy the $2a$ sites, which would have
increased $K_1$ (see $K_1$ values for $[2a]$ and $[2a,2a]$ in
Table~\ref{tab:H_a}).  However, the formation probability of the
configurations involving $12k$ site (such as $[12k]$, $[2a,12k]$ and
$[12k,12k]$) increases significantly as the temperature rises due to
the entropy contribution of the free energy.  At the annealing
temperature Al$^{3+}$ ions are much more likely to occupy the $12k$
site than the $2a$ site.  This causes the magnetocrystalline
anisotropy constant $K_1$ of Al-substituted SFO to be reduced with the
increase of Al fraction $x$, consistent with the experimental
measurement reported by Albanes \cite{Albanese1979}.  Despite of this,
$M_s$ is reduced more significantly than $K_1$ and this causes the
anisotropy field $H_a$ in Eq.~(\ref{eq:H_a}) to increase as the
concentration of Al $x$ is increased from 0 to 0.5 to 1.0 as shown in
Table~\ref{tab:H_a}.  This result is consistent with several
experimental results \cite{Ustinov:2009, Wang2012}.

\section{Conclusions}

Using the first-principles total energy calculations based on density
functional theory, we obtained the ground state structures and
associated formation probabilities at finite temperatures for
Al-substituted SFO, \SFAO\ with $x=0.5$ and $1.0$.  The structures
derived from our calculations show that the total magnetic moment of
the SFO unit cell is reduced as the fraction of Al atoms increases.
This reduction of magnetization is explained by the fact that the
non-magnetic Al atoms prefer to replace Fe$^{3+}$ ions at two of the
majority spin sites, $2a$ and $12k$, eliminating their positive
contribution to the total magnetization.  Our model also explains the
increase of the observed anisotropy field when the fraction of Al in
SFO is increased.  At the annealing temperature Al$^{3+}$ ions are
much more likely to occupy the $12k$ site than the $2a$ site.
Although this causes the magnetocrystalline anisotropy to decrease
slightly, the reduction in the saturation magnetization is larger and
their combined effect causes the magnetic anisotropy field of
Al-substituted SFO to be reduced with increase of Al fraction $x$. Our
results are consistent with the available experimental measurement on
Al-substituted SFO.

\section{Acknowledgments}

This work was supported in part by the U.S. Department of Energy
ARPA-E REACT program under Award Number DE-AR0000189 and the Center
for Computational Science (CCS) at Mississippi State
University. Computer time allocation has been provided by the High
Performance Computing Collaboratory (HPC$^2$) at Mississippi State
University.

\bibliographystyle{apsrev4-1}  
\bibliography{SFO}

\end{document}